\documentclass[preprint2]{aastex}
\usepackage{color}
\usepackage{graphicx}
\usepackage{bm}
\usepackage{amsmath}

\begin{document}
\title{Multi-dimensional radiative transfer to analyze Hanle effect in 
Ca {\sc ii} K line at 3933 \AA\,}
\author{L.~S.~Anusha$^{1,2}$ and K.~N.~Nagendra$^{1}$}
\affil{$^1$Indian Institute of Astrophysics, Koramangala,
2nd Block, Bangalore 560 034, India}
\affil{$^2$Max Planck Institute for Solar System Research, Katlenburg-Lindau,
37191, Germany}

\email{bhasari@mps.mpg.de,knn@iiap.res.in}

\begin{abstract}
Radiative transfer (RT) studies of the linearly polarized spectrum of the Sun
(the second solar spectrum) have generally focused on the line formation,
with an aim to understand the vertical structure of the solar atmosphere 
using one-dimensional (1D) model atmospheres. 
Modeling spatial structuring in the observations of the linearly polarized 
line profiles requires the solution of multi-dimensional 
(multi-D) polarized RT equation and a model solar atmosphere obtained by 
magneto-hydrodynamical (MHD) simulations of the solar atmosphere. 
Our aim in this paper is to analyze the chromospheric resonance line 
Ca {\sc ii} K at 3933 \AA\ using multi-D polarized RT with Hanle effect and 
partial frequency redistribution in line scattering. 
We use an atmosphere which is constructed by a two-dimensional snapshot 
of the three-dimensional MHD simulations of the solar photosphere, 
combined with columns of an 1D atmosphere in the chromosphere. This paper 
represents the first application of polarized multi-D RT to explore the 
chromospheric lines using multi-D MHD atmospheres, with PRD as the 
line scattering mechanism. We find that the horizontal inhomogeneities caused 
by MHD in the  lower layers of the atmosphere are responsible for strong spatial
inhomogeneities in the wings of the linear polarization profiles, 
while the use of horizontally homogeneous chromosphere (FALC) produces 
spatially homogeneous linear polarization in the line core. Introduction of 
different magnetic field configurations modifies the line core 
polarization through 
Hanle effect and can cause spatial inhomogeneities in the line core. 
A comparison of our theoretical profiles with the observations of this line 
shows that the MHD structuring in the photosphere is sufficient to 
reproduce the line wings and in the line core, only line center polarization
can be reproduced using Hanle effect. For a simultaneous modeling of the line 
wings and the line core (including the line center), MHD atmospheres with
inhomogeneities in the chromosphere are required.
\end{abstract}
\keywords{line: formation -- radiative transfer -- polarization --
scattering -- magnetic fields -- Sun: atmosphere}

\section{INTRODUCTION}
\label{intro}
Scattering polarization in strong resonance lines such as Ca {\sc i} 4227 \AA\,
and Ca {\sc ii} K at 3933 \AA\, can be used to diagnose the chromospheric 
magnetic fields \citep[see e.g.,][]{mf92,reneetal05,sametal09,anuetal10,
anuetal11b,hfetal12} and also to explore the temperature bifurcation in the 
chromosphere \citep{reneetal06,renejos07a,renejos07b}. 
All these studies consider one-dimensional 
(1D) models of the solar atmosphere such as FALC \citep{fontenla93}. However
the solar atmosphere is extremely inhomogeneous to be well represented by 1D
stratification of the physical quantities. Observations using ZIMPOL 
\citep{achimetal04} show strong spatial variation of the linear 
polarization along the slit, indicating spatial structuring in the atmosphere 
\citep[see e.g.,][]{biandaetal03,jos06,sametal09}. 
A natural explanation of this spatial structuring in the line core is in 
terms of Hanle effect from spatially varying magnetic fields. 
The spatial distribution of linear polarization, both in the core and the wings 
is caused by the local departures from axial symmetry of the atmosphere.
In the line core, this effect is entangled with the effect of spatially varying 
magnetic fields \citep[see][]{jos06}. To model such 
observations one has to solve multi-dimensional (multi-D) polarized radiative 
transfer (RT) equation with Hanle effect and partial frequency 
redistribution (PRD) in line scattering.
Methods have been developed to solve this problem in a series of papers 
\citep[][]{anuknn11a,anuetal11a,anuknn11b,anuknn11c,anuknn11d}.
In these papers we used isothermal atmospheres and hypothetical lines for the
sake of simplicity and to establish the relevant RT equation and its solution.
In the present paper we apply the methods developed in the above series of 
papers to the case of Ca {\sc ii} K line at 3933 \AA\,. 

Construction of MHD solar atmospheres achieved considerable success to 
represent the `solar photosphere' \citep[see e.g.,][and references cited 
therein]{nordstein91,vogleretal05}. Such photospheric models have been used to
analyze photospheric lines such as Sr {\sc i} 4607 \AA\, using polarized 
three-dimensional (3D) RT
equation and assuming complete frequency redistribution (CRD) 
\citep[see e.g.,][and references cited therein]
{jtbetal04,jtbshch07,shchjtb11}. However constructing MHD models that very well
represent the `solar chromosphere' remains still a challenging problem. 
A detailed review on the MHD and RT in 3D model atmospheres can be found 
in \citet{carl09}. 

In this paper we consider a 2D model, which is a combination of a vertical 
slice through the MHD snapshot from \citet[][]{nordstein91} and the 1D 
hydrostatic FALC model, replicated horizontally and joined smoothly in the 
vertical direction (provided by Han Uitenbroek, through a private 
communication). This atmosphere does not have any horizontal
inhomogeneities in the chromospheric layers. Hereafter we call this atmosphere
as MHD-FALC model atmosphere. This atmosphere may not quite well represent the
reality, but it certainly acts as an intermediate step between
the existing 1D models and the real multi-D MHD chromospheric atmospheres, 
the research on which is not established yet. Using this atmosphere we 
study the synthetic linear polarization profiles of Ca {\sc ii} K line at 
3933 \AA\,. 

In Section~\ref{formulation} we give all the governing equations of the
problem. In Section~\ref{method} we provide the mathematical background
of the solution method. In Section~\ref{numerics} we provide important
numerical details. Section~\ref{param} is devoted to a discussion on 
the spatial dependence of some crucial physical parameters. 
In Section~\ref{contrib} we discuss the
contribution functions. In Section~\ref{results} we discuss the results of our
studies. Section~\ref{conclusions} is devoted to the conclusions. In 
Appendix~\ref{prebicg} we give some mathematical details of the methods used
in this paper.

\section{FORMULATION OF THE PROBLEM}
\label{formulation}
For the line under consideration we first solve the unpolarized 
non-LTE multi-level RT equation and statistical equilibrium equation 
simultaneously and iteratively, using the code of \citet{hu00,hu01,hu06} (the 
RH-code). 
This code uses the Multi-level Accelerated Lambda Iteration (MALI) scheme 
of \citet{rh91,rh92}.
We consider 6 energy levels with 5 line transitions and 5 continuum transitions.
The main lines at 3933 \AA\, and 3968 \AA\, are treated in 
PRD and other lines in CRD. From the RH-code, we obtain level 
populations, radiative and collisional rates, opacities, and also the center 
to limb variation of the unpolarized intensity profiles. Keeping these 
quantities fixed, we then solve the polarized two-level RT equation 
for the main line under consideration. We use 
sophisticated domain based PRD theory formulated by 
\citet{bom97a,bom97b} (particularly the approximation level III). 
We start with the unpolarized source function as an initial solution, 
and use the Stabilized Preconditioned BiConjugate Gradient (Pre-BiCG-STAB)
technique \citep[see][]{anuetal09,anuetal11a,anuknn11b}. We work with the 
irreducible spherical tensor formulation of multi-D RT developed in 
\citet{anuknn11a, anuknn11b}. The 2D short-characteristics formal solution
is used \citep[see][]{auerfp94,kunauer88}.

\begin{figure*}
\centering
\includegraphics[scale=0.5]{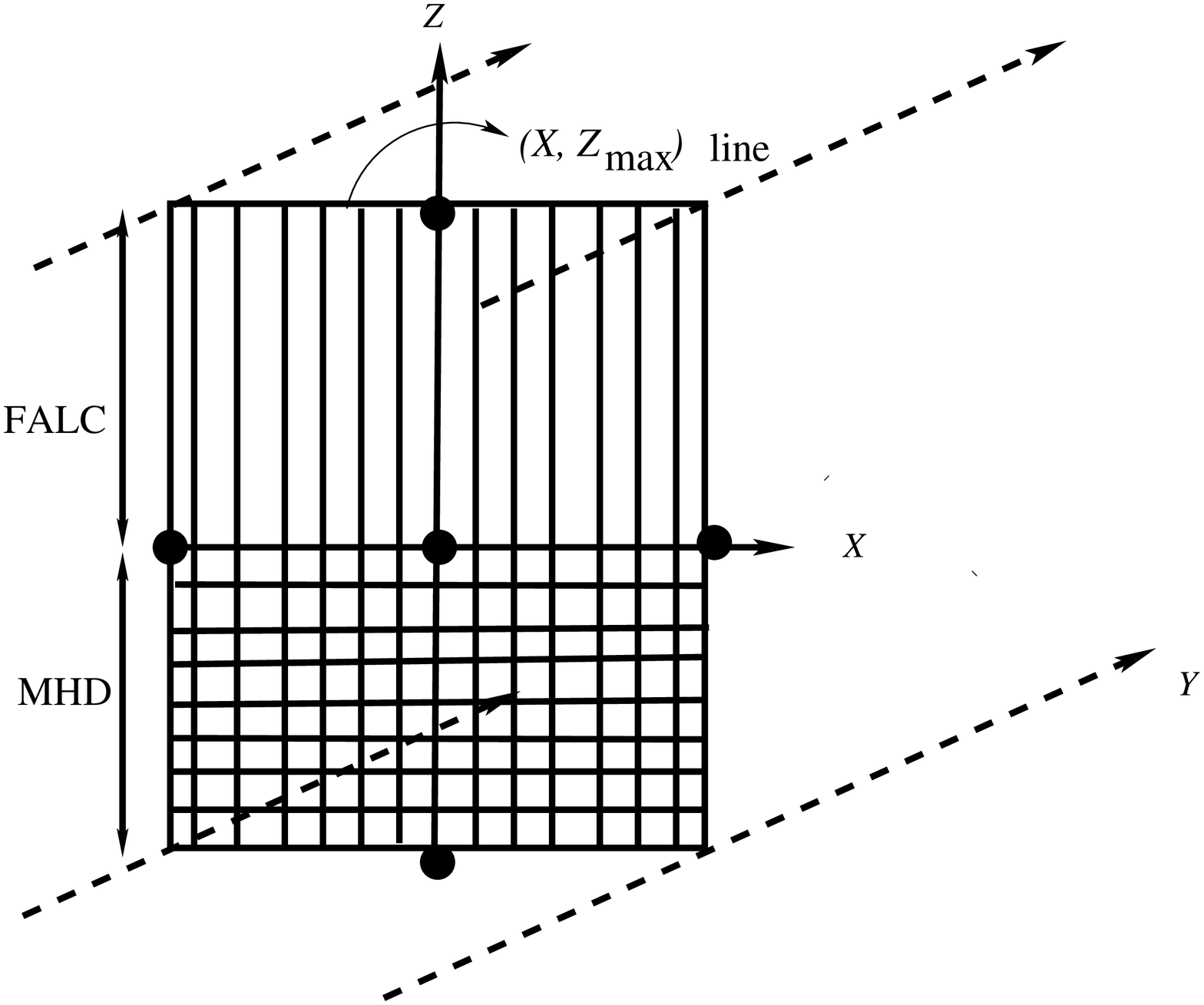}
\caption{This figure shows the geometry of the atmosphere
that we have used, namely a 2D cross section of a 3D
MHD atmosphere in the photosphere and columns of FALC atmosphere in the 
chromosphere. The emergent solutions on the line $(X, Z_{\rm max})$ 
marked here, are shown as images in the 
Figures~\ref{fig-iqu1}-\ref{fig-iqu4}.}
\label{fig-geometry}
\end{figure*}

The 3D MHD atmospheres are cuboids with all the
physical parameters depending on the spatial variables $x$, $y$ and $z$. 
In MHD-FALC model we consider a 2D snapshot along $Y$ direction so that in our 
calculations we have the physical parameters dependent only on $x$ and $z$ 
(see Figure~\ref{fig-geometry}).
We assume a periodic 2D medium with periodic horizontal boundary conditions.
For a ray traveling in the direction $\bm{\Omega}$ 
the 2D RT equation at a point $\bm{r}=(x,z)$ in the Cartesian co-ordinate 
system is given by
\begin{eqnarray}
&&\bm{\Omega}\cdot\nabla \bm{I}(\lambda, \bm{\Omega}, \bm{r})=
\nonumber \\
&&-[\kappa_{\rm tot}(\lambda, \bm{r})]
\left[\bm{I}(\lambda, \bm{\Omega}, \bm{r})- 
\bm{S}(\lambda, \bm{\Omega}, \bm{r})\right],
\label{rte}
\end{eqnarray}
where $\bm{I}$ = $(I,Q, U)^T$ is the Stokes vector and
$\bm{S}=(S_I, S_Q, S_U)^T$ is the source vector. 
The ray direction is defined by $\bm{\Omega}=(\theta, \varphi)$
where $\theta$ and $\varphi$ denote the inclination and the azimuth
of the scattered ray. The direction cosines of the ray are
$(\sin\theta\,\cos \varphi\,, \sin\theta\,\sin \varphi\,, \cos \theta)$
=$(\gamma,\eta,\mu)$.
The total opacity is
\begin{equation}
\kappa_{\rm tot}(\lambda, \bm{r})=[\kappa_l(\bm{r}) \phi(\lambda, \bm{r}) + 
\kappa_c(\lambda, \bm{r}) +\sigma_c(\lambda, \bm{r})].
\end{equation}
Here $\kappa_l$, $\kappa_c$ and $\sigma_c$
are wavelength averaged line opacity, continuum 
absorption and continuum scattering coefficients respectively. 
$\phi$ is the normalized Voigt profile function. The profile function
depends on $\bm{r}$ through the damping parameter $a$
which depends on the radiative de-excitation rate $\Gamma_R$, elastic and 
inelastic collision rates $\Gamma_E$ and $\Gamma_I$ respectively 
and the Doppler width $\Delta \nu_D$ so that
\begin{equation}
a=\frac{\Gamma_R+\Gamma_E+\Gamma_I}{4 \pi \Delta \nu_D}.
\label{gammatot}
\end{equation}
For the Ca {\sc ii} K line at 3933 \AA\, $\Gamma_R=1.5 \times 10^8 s^{-1}$.
$\Gamma_E$ is computed taking into account
the van der Waal's broadening (arising due to elastic collisions with neutral
hydrogen) and Stark broadening (arising due to interactions with free
electrons). $\Gamma_I$ includes the inelastic collision
processes like collisional de-excitation by electrons and protons,
collisional ionization by electrons, and charge exchange
processes \citep[see][]{hu01}.
Here $\Delta \nu_D = \sqrt{{2 k_BT}/{M_a}+v^2_{turb}}/\lambda_0$, 
with $k_B$ the Boltzmann constant, $T$ the temperature, $M_a$ the 
mass of the atom, $v_{turb}$ the micro-turbulent velocity (taken as 1 km/s), 
and $\lambda_0$ the line center wavelength.
In a two-level model atom with unpolarized ground level, the total
source vector $\bm{S}$ is defined as
\begin{eqnarray}
&&\bm{S}(\lambda, \bm{\Omega}, \bm{r})=\frac{\kappa_l(\bm{r})
\phi(\lambda, \bm{r}) \bm{S}_{l}(\lambda, \bm{\Omega}, \bm{r})}
{\kappa_{\rm tot}(\lambda, \bm{r})}
\nonumber \\ 
&&+\frac{\sigma_c(\lambda, \bm{r}) \bm{S}_{c}(\lambda, \bm{\Omega}, \bm{r}) 
+ \kappa_c(\lambda, \bm{r}) {B}_{\lambda}(\bm{r}){\bm U}}
{\kappa_{\rm tot}(\lambda, \bm{r})}. 
\nonumber \\
\label{s-tot}
\end{eqnarray}
Here $\bm{U}=(1,0,0)^T$ and $B_{\lambda}$
is the Planck function. The line source vector is
\begin{eqnarray}
&&{\bm{S}}_{l}(\lambda, \bm{\Omega}, \bm{r})=
\epsilon{B}_{\lambda}(\bm{r})\bm{U}\nonumber \\
&&\!\!\!\!\!\!\!\!\!\!+\int_{-\infty}^{+\infty}\oint
\frac{\hat{R}(\lambda, \lambda', \bm{\Omega}, \bm{\Omega}', \bm{r}, \bm{B})}
{\phi(x)}
\nonumber \\
&& \times\, {\bm{I}}(\lambda', \bm{\Omega}', \bm{r})
\,\frac{d\bm{\Omega}'}{4\pi}\,d\lambda'.
\label{sl}
\end{eqnarray}
Here $\hat{R}$ is the Hanle redistribution matrix \citep[][]{bom97a,bom97b}.
$\bm{B}$ is the vector magnetic field taken to be a free parameter.
The continuum scattering source vector is
\begin{equation}
{\bm{S}}_{c}(\lambda, \bm{\Omega}, \bm{r})=
\oint \hat{P}(\bm{\Omega}, \bm{\Omega}')
{\bm{I}}(\lambda, \bm{\Omega}', \bm{r})\,\frac{d\bm{\Omega}'}{4\pi},
\label{sc}
\end{equation}
where $\hat{P}$ is the Rayleigh scattering phase matrix
\citep[][]{chandra60}. For simplicity, frequency coherent
scattering is assumed for the continuum.
The thermalization parameter $\epsilon$ is defined by
$\epsilon={\Gamma_I}/(\Gamma_R+\Gamma_I)$.
($\lambda'$, $\bm{\Omega}'$) and
($\lambda$, $\bm{\Omega}$) in Equation~(\ref{sl}) refer to the wavelength
and direction of the incoming and the outgoing rays, respectively.

\section{METHOD OF SOLUTION}
\label{method}
In \citet{anuknn11a,anuknn11b} we describe a method of decomposing
the Stokes vectors and Stokes source vectors into irreducible
spherical tensors ${\mathcal T}^K_Q$ \citep[see e.g.,][]{ll04} which transforms 
the multi-D RT equation into a simpler form that can be solved using
any iterative method (see Section~\ref{ikq}). In \citet{anuetal11a} and
\citet{anuknn11b} we develop a fast  iterative method called the 
Pre-BiCG-STAB for polarized multi-D RT with PRD. We use both these techniques 
here to solve the relevant RT equation. Because of the 
use of continuum scattering the algorithm for the Pre-BiCG-STAB should be 
modified slightly, which is described in Appendix~\ref{prebicg}.

\subsection{Irreducible spherical tensors for multi-D RT}
\label{ikq}
Using irreducible spherical tensors we can represent the linearly polarized 
radiation field through a six component vector denoted by 
$\bm{\mathcal I}=(I^0_0, I^2_0, I^{2,x}_1, I^{2,y}_1, I^{2,x}_2, I^{2,y}_2)^T$
which are related to the Stokes parameters $I,Q$ and $U$ through the following
expressions \citep[see][]{hf07}.
\begin{eqnarray}
&&I(\lambda, \bm{\Omega}, \bm{r}) = I^0_0 +
\frac{1}{2 \sqrt{2}} (3 \cos^2\theta -1) I^2_0 \nonumber \\
&&-\sqrt{3} \cos \theta \sin \theta (I^{2,x}_1 
\cos \varphi-I^{2,y}_1 \sin \varphi) \nonumber \\ 
&&+ \frac{\sqrt{3}}{2} (1-\cos^2\theta)
(I^{2,x}_2 \cos 2 \varphi-I^{2,y}_2 \sin 2 \varphi), \nonumber \\
\label{transform-1}
\end{eqnarray}
\begin{eqnarray}
&&Q(\lambda, \bm{\Omega}, \bm{r})= -\frac{3}{2 \sqrt{2}} 
(1- \cos^2\theta) I^2_0 \nonumber \\
&&-\sqrt{3} \cos \theta \sin \theta (I^{2,x}_1 
\cos \varphi-I^{2,y}_1 \sin \varphi) \nonumber \\ 
&&-\frac{\sqrt{3}}{2} (1+\cos^2\theta)
(I^{2,x}_2 \cos 2 \varphi-I^{2,y}_2 \sin 2 \varphi),\nonumber \\
\label{transform-2}
\end{eqnarray}
\begin{eqnarray}
&&U(\lambda, \bm{\Omega}, \bm{r}) = \sqrt{3} \sin \theta
(I^{2,x}_1 \sin \varphi+I^{2,y}_1 \cos \varphi) \nonumber \\ 
&&+ \sqrt{3} \cos \theta 
(I^{2,x}_2 \sin 2 \varphi+I^{2,y}_2 \cos 2 \varphi).
\label{transform-3}
\end{eqnarray}
Here $I^0_0$, $I^2_0$, $I^{2,x}_1$, $I^{2,y}_1$, $I^{2,x}_2$ and $I^{2,y}_2$
are components of the vector $\bm{\mathcal I}$ all of which depend on the
parameters $\lambda$, $\bm{\Omega}$, and $\bm{r}=(x,y,z)$.

Together with a six dimensional source vector denoted by 
$\bm{\mathcal S}=(S^0_0, S^2_0, S^{2,x}_1, S^{2,y}_1, S^{2,x}_2, 
S^{2,y}_2)^T$ the irreducible Stokes vector $\bm{\mathcal I}$ 
satisfies an RT equation given by
\begin{eqnarray}
&&-\frac{1}{\kappa_{\rm tot}(\lambda, \bm{r})}\bm{\Omega} \cdot
\bm{\nabla}\bm{\mathcal{I}}(\lambda, \bm{\Omega}, \bm{r}) = \nonumber \\
\!\!\!\!\!\!&&[\bm{\mathcal{I}}(\lambda, \bm{\Omega}, \bm{r})-
\bm{\mathcal S}(\lambda,\bm{r})].
\label{rte-reduced}
\end{eqnarray}

For a two-level atom model with unpolarized ground level 
$\bm{\mathcal S}(\lambda, \bm{r})$ takes the form
\begin{equation}
\bm{\mathcal S}(\lambda, \bm{r})=p_l \bm{\mathcal S}_{l}(\lambda,\bm{r})
+ p_c\bm{\mathcal S}_c(\lambda,\bm{r}) + p_a {B}_{\lambda}(\bm{r})\bm{\mathcal U},
\label{stot-reduced}
\end{equation}
with
\begin{eqnarray}
&&p_l=\kappa_l(\bm{r}) \phi(\lambda,\bm{r}) / \kappa_{\rm tot}(\lambda,\bm{r}),
\nonumber\\
&&p_c=\sigma(\lambda,\bm{r}) / \kappa_{\rm tot}(\lambda,\bm{r}),
\nonumber\\
&&p_a=\kappa_c(\lambda,\bm{r}) / \kappa_{\rm tot}(\lambda,\bm{r}),
\label{px}
\end{eqnarray}
and $\bm{\mathcal U}=(1,0,0,0,0,0)^T$.
The irreducible line source vector $\bm{\mathcal S}_{l}$ is 
\begin{eqnarray}
\bm{\mathcal S}_{l}(\lambda, \bm{r})=
\bm{\mathcal G}(\lambda,\bm{r})+\bm{\mathcal{J}}(\lambda, \bm{r}),
\label{sl-reduced}
\end{eqnarray}
where
\begin{eqnarray}
&&\bm{\mathcal{J}}(\lambda, \bm{r})=\frac{1}{\phi(\lambda,\bm{r})} 
\int_{-\infty}^{+\infty} d\lambda' \nonumber \\
\!\!\!\!\!\!&& \times
\oint\frac{d\bm{\Omega}'} {4 \pi} 
\hat{W}\Big\{\hat{M}_{\rm II}(\bm{B},\lambda,\lambda')
r_{\rm II}(\lambda,\lambda') \nonumber \\
\!\!\!\!\!\!&&+\hat{M}_{\rm III}(\bm{B},\lambda,\lambda')
r_{\rm III}(\lambda, \lambda') \Big\}
\hat{\Psi}(\bm{\Omega}') \nonumber \\
&&\times \bm{\mathcal{I}}(\lambda',\bm{\Omega}',\bm{r}),
\label{jbar}
\end{eqnarray}
with $\bm{\mathcal G}(\lambda,\bm{r})=
\epsilon {B}_{\lambda}(\bm{r})\bm{\mathcal U}$. 
The irreducible continuum scattering source vector is
\begin{equation}
{\bm{\mathcal S}}_{c}(\lambda, \bm{r})=
\oint \hat{\Psi}(\bm{\Omega}') \bm{\mathcal{I}}(\lambda,\bm{\Omega}',\bm{r})
\,\frac{d\bm{\Omega}'}{4\pi}.
\label{sc-reduced}
\end{equation}
The matrix $\hat{\Psi}$ represents the reduced phase matrix for
the Rayleigh scattering. Its elements are listed in Appendix D
of \citet{anuknn11b}. The elements of the
matrices $\hat{M}_{\rm II,III}(\bm{B},\lambda,\lambda')$ for the Hanle
effect are derived in \citet{bom97a,bom97b}. The dependence of the matrices
$\hat{M}_{\rm II,III}(\bm{B},\lambda,\lambda')$ on $\lambda$ and $\lambda'$ 
is related to the definitions of the frequency domains \citep[see approximation
level III of][]{bom97b}. The functions $r_{\rm II}$ and $r_{\rm III}$ are the
angle-averaged PRD functions of \citet[][]{hum62}.
$\hat{W}$ is a diagonal matrix written as
\begin{equation}
\hat{W}=\textrm{diag}\{W_0,W_2,W_2,W_2,W_2,W_2\}.
\label{w}
\end{equation}
Here the weight $W_0=1$ and the weight $W_2$ depends on the angular
momentum quantum number of the line under consideration 
\citep[see][]{ll04}. For the Ca {\sc ii} K line at 3933 \AA\, $W_2=1$.

We use the short characteristics method \citep[][]{kunauer88,auerfp94} 
for computing the formal solution of Equation~(\ref{rte-reduced}). 
Let ${\rm MOP}$ be a segment of a ray, also called as a short characteristics 
stencil. The irreducible Stokes vector $\bm{\mathcal{I}}$ at ${\rm O}$ is given by
\begin{eqnarray}
&&\bm{\mathcal{I}}_{\rm O}((\lambda, \bm{\Omega}, \bm{r})=
\bm{\mathcal{I}}_{\rm M}(\lambda, \bm{\Omega}, \bm{r}) \exp[-\Delta \tau_{\rm M}]
\nonumber \\
&&+{\psi}_{\rm M}(\lambda, \bm{\Omega}, \bm{r})
\bm{\mathcal{S}}_{\rm M}(\lambda, \bm{r})\nonumber \\
&&+{\psi}_{\rm O}(\lambda, \bm{\Omega}, \bm{r})
\bm{\mathcal{S}}_{\rm O}(\lambda, \bm{r})\nonumber \\
&&+{\psi}_{\rm P}(\lambda, \bm{\Omega}, \bm{r})
\bm{\mathcal{S}}_{\rm P}(\lambda, \bm{r}),\nonumber \\ 
\label{int_sc}
\end{eqnarray}
where $\bm{\mathcal{S}}_{\rm M,O,P}$ are the irreducible source vectors at
${\rm M}$, ${\rm O}$ and ${\rm P}$.
The quantity $\bm{\mathcal{I}}_{\rm M}$ is the upwind irreducible Stokes
vector at the point ${\rm O}$. The coefficients ${\psi}$
depend on the optical depth increments in $X$ and $Z$ directions
and are given in \citet{auerfp94}.

\subsection{Numerical details}
\label{numerics}
In this paper we have developed a 2D polarized RT code to compute 
Stokes profiles in 2D snapshots of an MHD atmosphere. 
This code treats PRD as the line scattering mechanism for a two-level 
atom system. The outputs from the RH-code are used as input data 
in the polarized RT 
computations (see Section~\ref{formulation} above). 
For our calculations we use equidistant spatial grids that are obtained 
from the model atmosphere, which is then converted into logarithmic $\tau$ 
grids. For frequency (or wavelength) grid we use Simpson quadrature. 
For angle integration we use Carlsson type A4 set which considers three
directions per octant with a total of twenty four
directions for the full atmosphere.

\section{SPATIAL VARIATION OF THE PHYSICAL PARAMETERS}
\label{param}
\begin{figure*}
\centering
\includegraphics[scale=0.4]{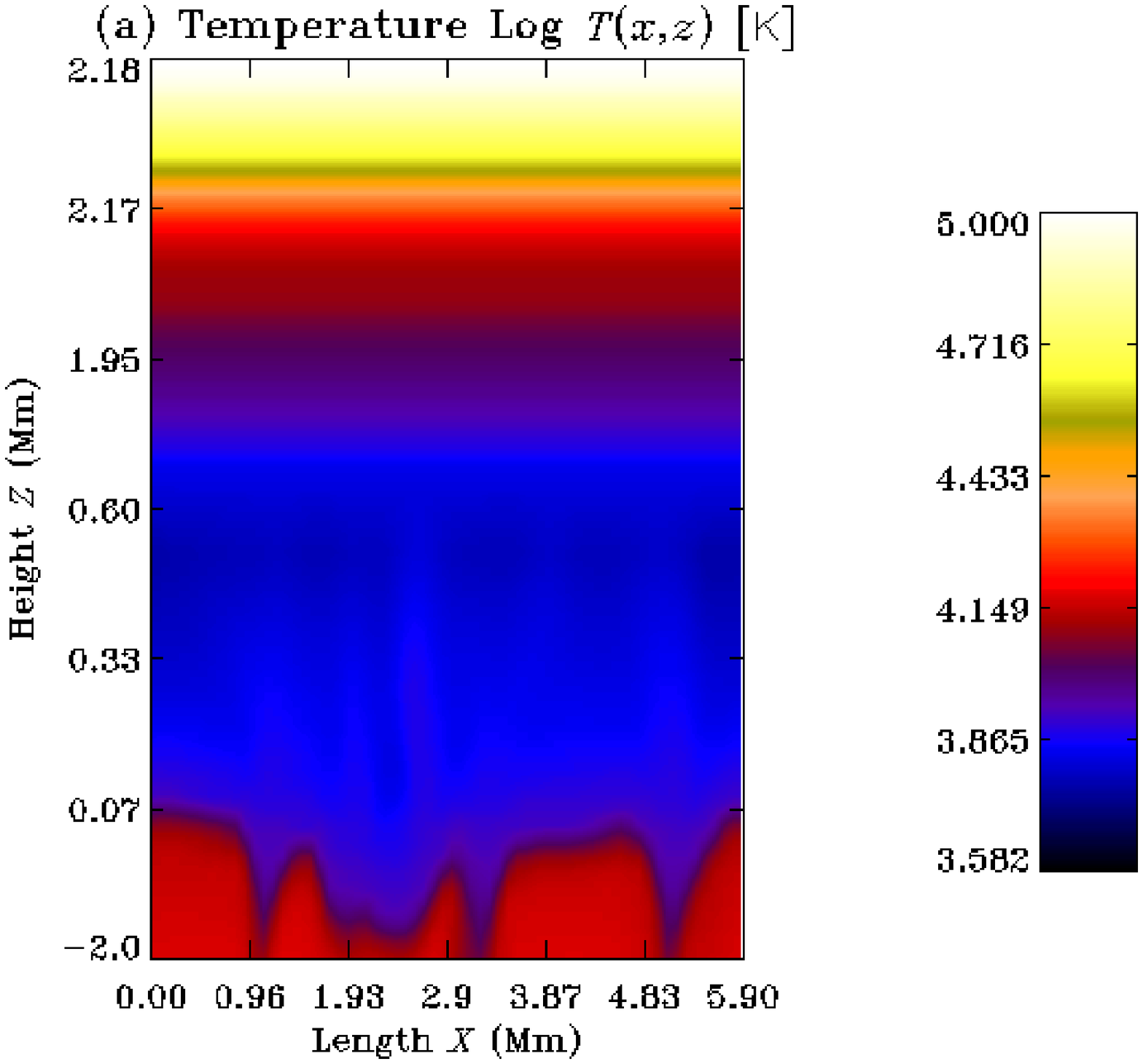}
\includegraphics[scale=0.4]{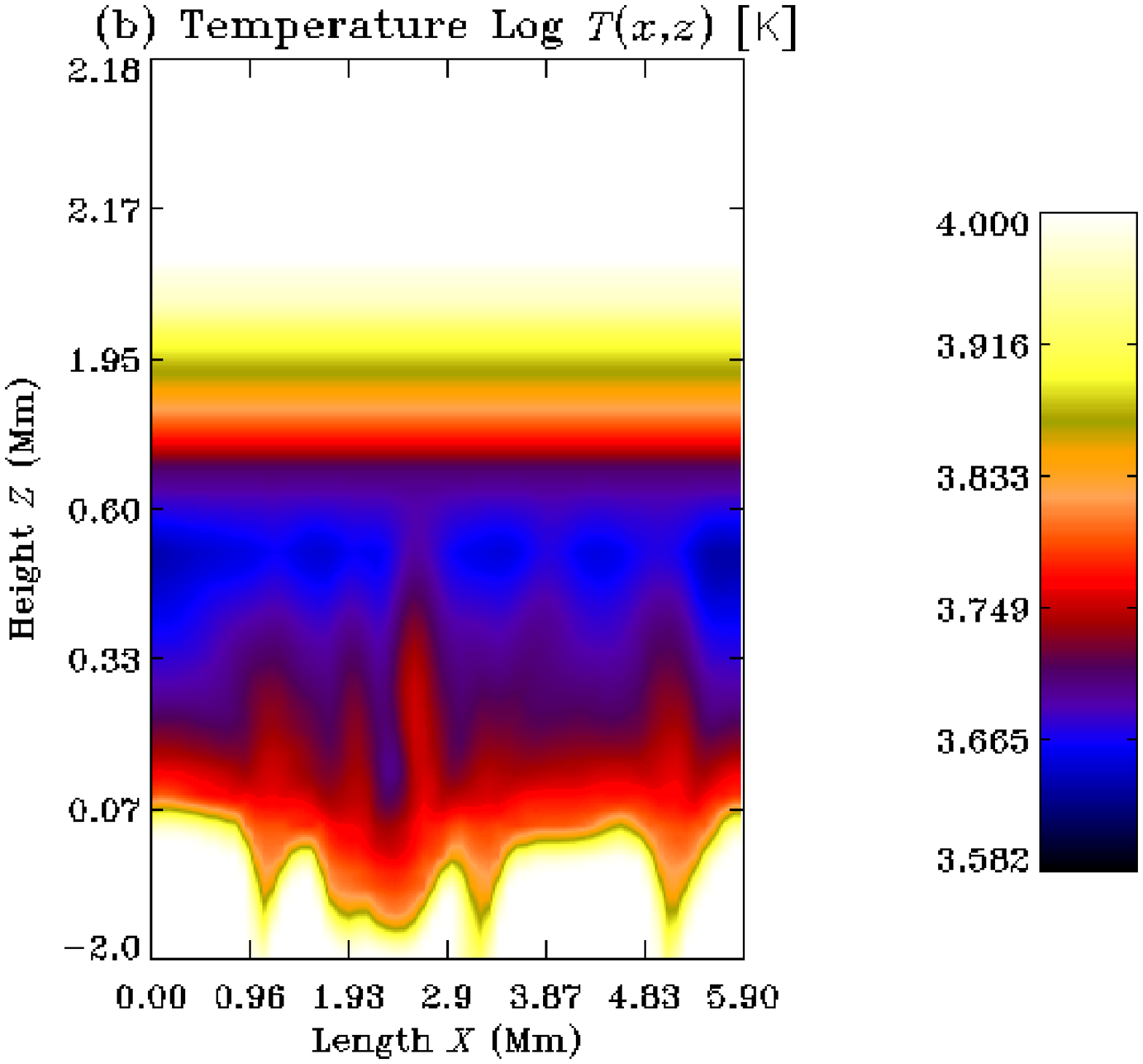}
\caption{Temperature structure $T$ (in log scale) in the MHD-FALC atmosphere. 
Panel (b) shows the variation in a smaller range of temperature $T$ 
between 3819 K and 10000 K.}
\label{fig-temp}
\end{figure*}

\begin{figure*}
\centering
\includegraphics[scale=0.4]{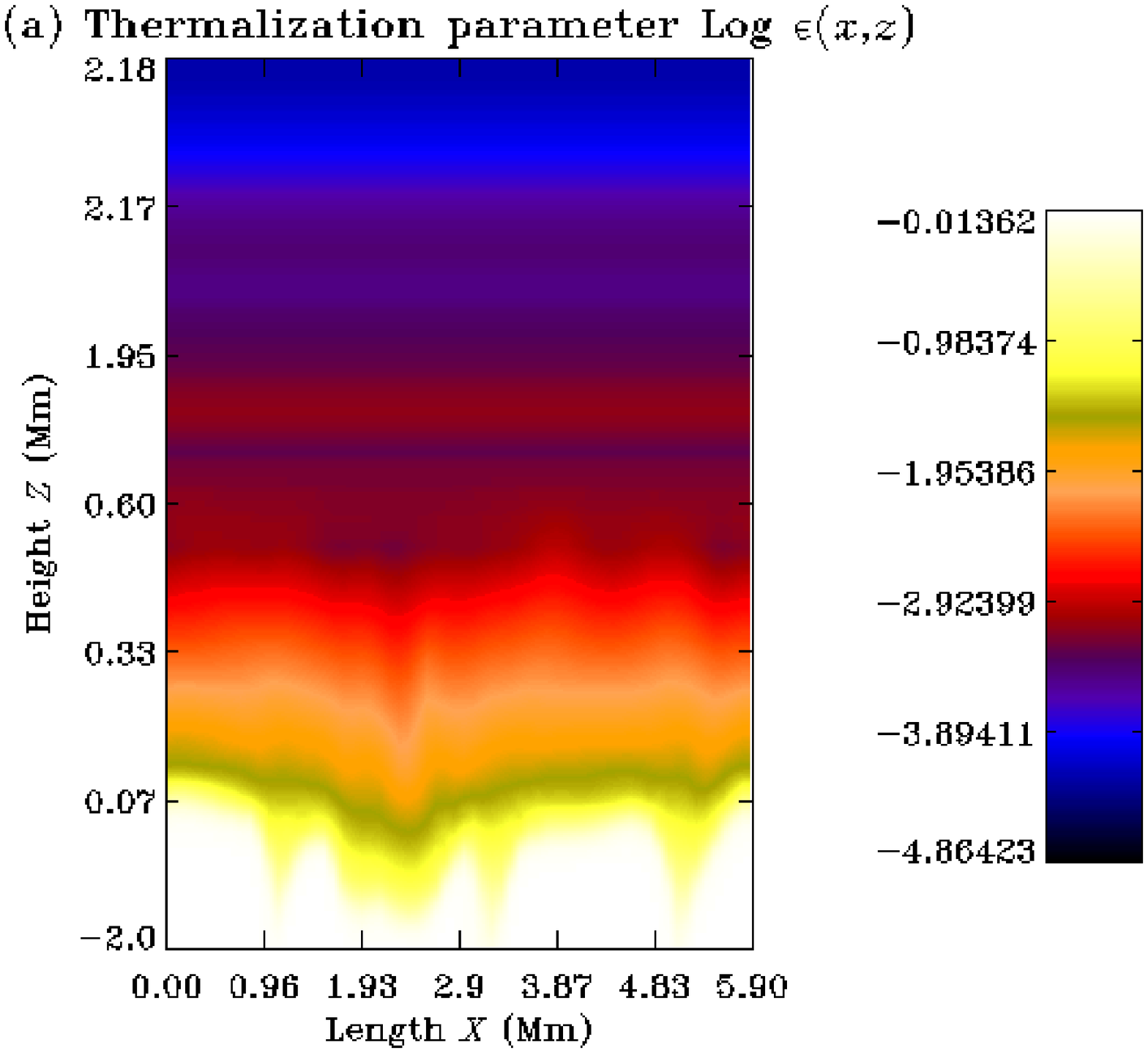}
\includegraphics[scale=0.4]{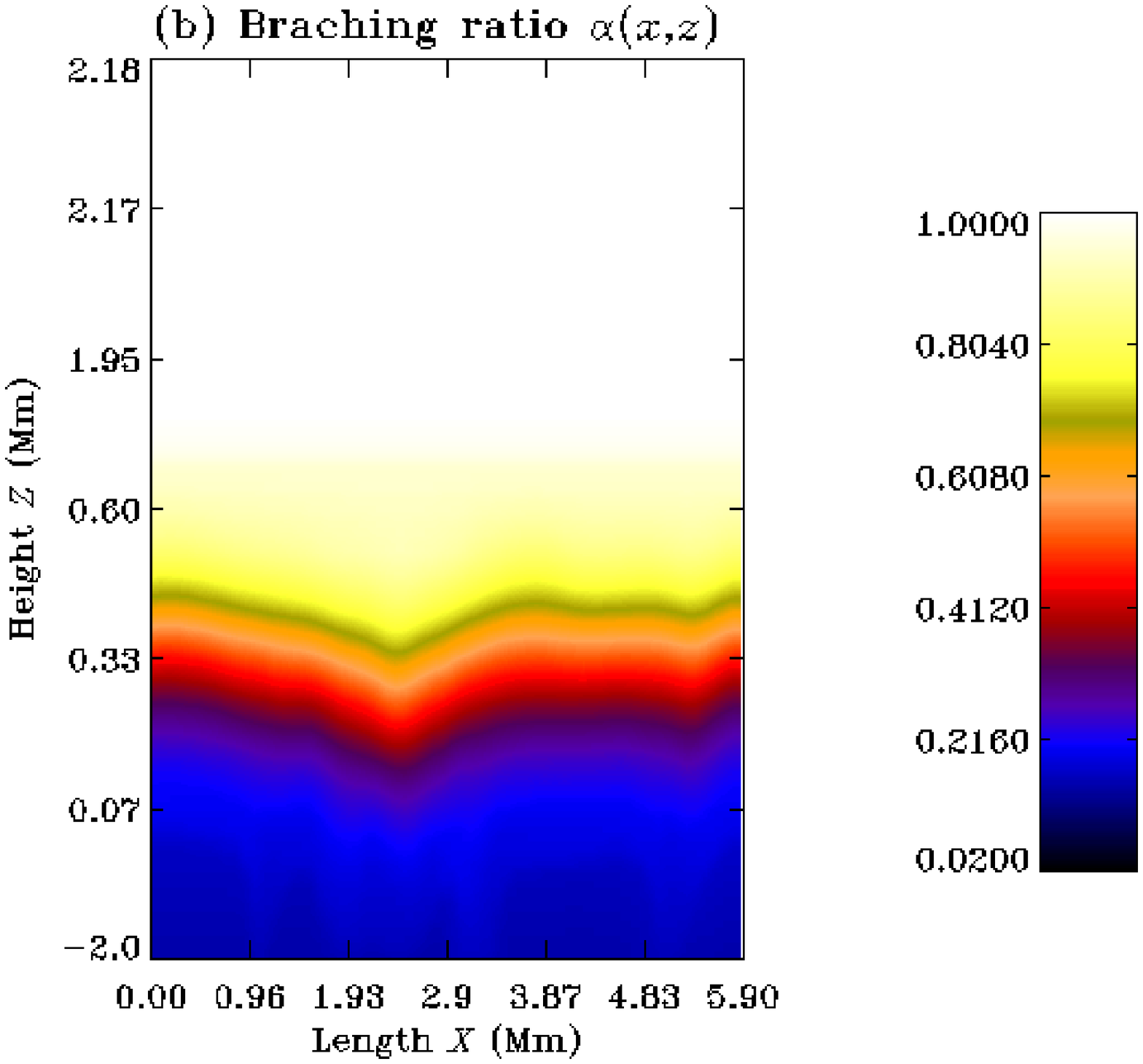}
\caption{The spatial variation of thermalization parameter $\epsilon$ (in 
log scale) and the branching ratio $\alpha$ (that appears in the PRD 
matrices) in the MHD-FALC atmosphere.}
\label{fig-epsilon-alpha}
\end{figure*}

In this section we focus our attention on the spatial variation of some of the 
crucial physical parameters on which the variation of the linearly polarized 
profiles sensitively depend. 
In Figures~\ref{fig-temp}(a) and (b) we show the structure of temperature 
(in log scale) in the MHD-FALC atmosphere along $X$ and $Z$ directions. 
In Figures~\ref{fig-epsilon-alpha}(a) and (b) we show respectively the 
thermalization parameter $\epsilon$ (in log scale) and 
the branching ratio $\alpha=\Gamma_R/(\Gamma_R+\Gamma_E+\Gamma_I)$ that 
multiplies $r_{\rm II}$ function in the PRD matrices \citep[][]{bom97b}.
Color bars in the respective figures represent the range of variation of the 
values of the physical parameters shown in corresponding figures.

In Figure~\ref{fig-temp}(a) we show the temperature structure Log $T$ 
in the MHD-FALC atmosphere, with $T$ in the range 3819 K to 10$^5$ K. 
Since the variation spans a wide range of values, the image looks nearly the 
same in the heights between 0.07 Mm to 0.65 Mm. To show the actual temperature 
variation in these heights we restrict the upper limit of temperature to 
10$^4$ K in Figure~\ref{fig-temp}(b), where one can notice a significant 
horizontal and vertical variation of temperature.
Above $\sim$ 0.65 Mm there is no horizontal inhomogeneity in the MHD-FALC 
atmosphere and therefore we see the same temperature in the horizontal 
direction. The vertical variation in these layers is the same as the well
known temperature variation in 1D FALC atmosphere \citep[see e.g., 
Figure 2 in][]{anuetal10}. 

The spatial variations of $\epsilon$ and $\alpha$ are more interesting for 
the line formation process. In Figure~\ref{fig-epsilon-alpha}(a) 
$\epsilon$ approaches unity in the deeper layers which means that the 
source function has dominant contribution from thermal sources in 
these layers. A gradual decrease in the value of $\epsilon$ with increase 
in height shows the increased contribution from scattering sources in 
those layers. $\epsilon$ shows some horizontal variations in the layers 
below 0.65 Mm. The parameter $\alpha$ shown in 
Figure~\ref{fig-epsilon-alpha}(b) multiplies the $r_{\rm II}$ function in the 
redistribution matrices and it represents the probability of frequency coherent 
scattering (in the atom's rest frame) in the medium. It approaches unity 
in the layers above 0.65 Mm showing that 
contribution from $r_{\rm II}$ dominates 
over that from $r_{\rm III}$ in these layers. In between 0.2 Mm to 0.65 Mm, 
$\alpha$ takes values between 0.3 -- 0.9 
which means that $r_{\rm II}$ continues to 
have a significant contribution to the line formation in these layers.
In the deepest layers where collisions dominate over scattering, $\alpha$ 
takes smaller values ($\sim$ 10$^{-2}$). In these layers contribution from 
$r_{\rm III}$ redistribution dominates over $r_{\rm II}$.

\section{CONTRIBUTION FUNCTIONS}
\label{contrib}
In this section we discuss the contribution functions in the context of 2D RT
and the corresponding heights of formation of the Ca {\sc ii} K line at 
different wavelength points.

Following \citet[][]{magain86} the contribution function $C_I$ for the specific 
intensity in the context of 2D RT may be defined as
\begin{eqnarray}
&& C_I(\lambda,\bm{r},\bm{\Omega})=\frac{1}{c_s(\bm{\Omega})} 
\ln(10)\frac{\kappa_{\rm tot}(\lambda,\bm{r})}
{\kappa_{\rm tot}(\lambda_0,\bm{r})}\nonumber\\
&&s(\lambda_0,\bm{r},\bm{\Omega})
S(\lambda,\bm{r}) \exp\left[{-s(\lambda,\bm{r},\bm{\Omega})}\right].
\label{c_i}
\end{eqnarray}
Here $S(\lambda,\bm{r})$ is the unpolarized source function and 
$s(\lambda,\bm{r},\bm{\Omega})$ is the photon path length defined
as 
\begin{eqnarray}
&&s(\lambda,\bm{r},\bm{\Omega})=\int_{\textrm{path}} 
ds(\lambda,\bm{r},\bm{\Omega}),\quad{\textrm{with}}\nonumber\\
&&ds(\lambda,\bm{r},\bm{\Omega})=\kappa_{\rm tot}(\lambda,\bm{r})
d{\rm x}/c_s(\bm{\Omega}) \quad {\textrm{or}} \nonumber\\
&&=\kappa_{\rm tot}(\lambda,\bm{r})d{\rm z}/c_s(\bm{\Omega}),
\end{eqnarray}
where the direction cosine $c_s(\bm{\Omega})=\gamma$ if the ray hits the 
horizontal $X$ axis and $c_s(\bm{\Omega})=\mu$ if the ray hits the vertical
$Z$ axis when it passes through the 2D medium \citep[see][]{kunauer88}.

The spatially averaged contribution functions $<C_I>$ are obtained by 
performing an arithmetic averaging of $C_I$ over the spatial $X$ grid. 
In Figure~\ref{fig-contrib} we plot $<C_I>$ for 5 selected 
wavelengths and two ray directions chosen to represent a near-limb and a 
near-disk-center lines of sight. In Table~\ref{table_1} these 5 selected 
wavelengths and the height at which $<C_I>$ reaches its maximum value are 
tabulated, which can be considered to be an approximate line formation 
height at that wavelength.
\begin{table*}
\begin{center}
\caption{The line formation heights (in Mm) at different wavelengths for
near-limb and near-disk-center lines of sight.}
\vspace{0.6cm}
\label{table_1}
\begin{tabular}{crrrrr}
\tableline\tableline
$\lambda$ &  Near-limb& Near-disk-center  \\
\tableline
3928.15 \AA\, & 0.25 Mm & 0.2 Mm\\
\tableline
3933.09 \AA\, & 0.65 Mm & 0.5 Mm\\
\tableline
3933.50 \AA\, & 2.15 Mm & 1.27 Mm\\
\tableline
3933.65 \AA\, & 2.16 Mm & 2.16 Mm\\
\tableline
3933.80 \AA\, & 2.16 Mm & 1.17 Mm\\
\tableline
\tableline
\end{tabular}
\end{center}
\end{table*}

For both the directions considered here, the far wings ($\sim$ 3928.15 \AA\,) 
and the near wings ($\sim$ 3933.09 \AA\,) are formed at or 
below 0.65 Mm. For the near-limb line of sight, the line core 
region (3933.50 \AA\,--3934 \AA\,) including blue core minimum ($\sim$ 
3933.50 \AA\,), red core minimum ($\sim$ 3933.80 \AA\,) and the line center 
($\sim$ 3933.65 \AA\,) are formed higher up in the 
atmosphere between 2 Mm -- 2.5 Mm. For the near-disk-center line of sight line
center is formed between 2 Mm to 2.5 Mm, but the core minima are formed just
above 1 Mm. The heights of formation are very useful to understand the 
spatial distribution of the linear polarization discussed in 
Section~\ref{results}.

\section{RESULTS AND DISCUSSIONS}
\label{results}
In this section we discuss the results of our investigations on the 
nature of the Stokes profiles calculated using our 2D polarized RT 
code with PRD as the scattering mechanism. The focus is on (1) the 
spatial variations of the Stokes profiles and 
(2) comparison of the observed and synthetic Stokes profiles computed
from our code. 

\subsection{Spatial variation of the Stokes profiles}
\label{r1}
It is well known that the strong resonance lines with broad
wings have their origin in the PRD coherent scattering mechanism 
($r_{\rm II}$ redistribution).
When combined with multi-D RT, the PRD effects manifest themselves 
also through enhanced inhomogeneous spatial distribution of the 
linear polarization profiles in the line wings \citep[see e.g.,][]{anuknn11b}. 
In that paper we consider an isothermal atmosphere. Temperature structure 
of the realistic atmosphere further enhances the inhomogeneities. 
These characteristics can be clearly seen in 
Figures~\ref{fig-iqu1}--\ref{fig-iqu4} which we discuss below.

In Figures~\ref{fig-iqu1}--\ref{fig-iqu4} we show the spatial variations of the
emergent synthetic spectra of Ca {\sc ii} K line at 3933 \AA\,
computed using MHD-FALC atmosphere.
We present the images of the $(I/I_c, Q/I, U/I)$ with the $\lambda$-scale on
the abscissae (covering the range 3913 \AA\,--3948 \AA\,) and the $X$-scale
on the ordinates. The images cover a range of values between minima and
maxima of the Stokes profiles. We have considered four directions
namely $(\mu,\varphi)$=$(0.3,160^{\circ})$, $(0.3,200^{\circ})$, 
$(0.8,135^{\circ})$ and $(0.8,225^{\circ})$. The two chosen $\mu$ values
represent near-limb ($\mu=0.3$) and near-disk-center ($\mu=0.8$) 
lines of sight. We chose two magnetic field ($\bm{B}$) configurations given by 
$(B,\theta_B,\chi_B)$ $=(20\,{\textrm G\,},45^{\circ},225^{\circ})$ and 
$(20\,{\textrm G\,},75^{\circ},225^{\circ})$. We assume the magnetic field 
$\bm{B}$ to be independent of spatial co-ordinates $X$ and $Z$. 
For clarity, the line core region is magnified in the middle 
panels in Figures~\ref{fig-iqu1}--\ref{fig-iqu4}. The wavelength range 
in these panels is 3932.9 \AA\, to 3934.2 \AA\,.
For the sake of discussions we refer to the spectrum in the wavelength range
3933.5 \AA--3934 \AA\, as the line core and those in the range
3913 \AA--3933.5 \AA\, and 3934 \AA\,--3948 \AA\, as the line wings.

\subsubsection{General characteristics}
\label{r11}
First we remark that due to the periodicity of the medium in the horizontal
direction considered for the calculations, the spatial distribution 
of $I/I_c$, $Q/I$ and $U/I$ is also periodic with respect to the horizontal
$X$ direction. As discussed in Section~\ref{contrib} and Table~\ref{table_1}
the Ca {\sc ii} K line wings are formed at or below a height of 
$\sim$ 0.65 Mm, and the line core is formed above this height. We note that 
0.65 Mm is the approximate height at which the MHD-FALC
atmosphere changes from 2D-MHD variation to horizontally homogeneous 
stratification represented by the 1D FALC atmosphere. Therefore, the
spatial structuring in the lower atmosphere causes significant spatial 
inhomogenity in the wings of $(I/I_c,Q/I,U/I)$ and spatial homogeneity 
(in the horizontal direction) in the higher layers of the atmosphere causes 
spatial homogeneity in the line core.
This is true for all the lines of sight in Figures~\ref{fig-iqu1}--
\ref{fig-iqu4}.

\subsubsection{Intensity distribution}
\label{r12}
A comparison of the spatial distribution of $I/I_c$ in 
Figures~\ref{fig-iqu1}--\ref{fig-iqu4} show that the emergent intensity 
is not as sensitive to the
radiation azimuth $(\varphi)$ as it is to the co-latitude $\theta$. In other
words the center to limb variation of $I/I_c$ is stronger than the 
axial-asymmetry of the emergent radiation field. In the far wings $I/I_c$ 
approaches unity, clearly because it is normalized to the nearby continuum 
value.

\subsubsection{$Q/I$ and $U/I$ in the line wings}
\label{r13}
Spatial distribution of $Q/I$ and $U/I$ for near-limb and near-disk-center
lines of sight have considerable differences (compare Figure~\ref{fig-iqu1}
with \ref{fig-iqu3} and \ref{fig-iqu2} with \ref{fig-iqu4}). For example, in
Figures~\ref{fig-iqu1}  and \ref{fig-iqu3}, $|Q/I|_{\rm max}$ amplitudes 
decrease from 2.5 \% to 0.36 \% from near-limb to the near-disk-center, and 
$|U/I|_{\rm max}$ amplitudes increase from 0.24 \% to 0.40 \%.
The distribution appears in the form of several domains within which $Q/I$ and
$U/I$ values are slowly varying.

The $Q/I$ and $U/I$ distributions in the wings respectively show a symmetry and
antisymmetry with respect to the radiation azimuth belonging to two opposing 
octants about the symmetry axis, for a given value of $\mu$ 
(compare Figure~\ref{fig-iqu1} 
with \ref{fig-iqu2} and \ref{fig-iqu3} with \ref{fig-iqu4}). This is a 
behavior of the Rayleigh scattered polarized radiation field in a 2D medium 
with respect to the symmetry axis \citep[see Appendix B of]
[for a proof]{anuetal11a}. Since Hanle scattering affects only the line core, 
and the Rayleigh scattering limit is recovered in the wings, we see the 
symmetries discussed above, only in the wings of these images. We note 
here that the azimuths in Figures~\ref{fig-iqu1} and \ref{fig-iqu3} lie 
on one side of the symmetry axis ($160^{\circ}$ and $135^{\circ}$ 
respectively) and those in Figures~\ref{fig-iqu2} and \ref{fig-iqu4} 
lie on the other side ($200^{\circ}$ and $225^{\circ}$ respectively) because 
of which symmetry relations become valid. Exact symmetry 
anti-symmetry in the values of $Q/I$ and $U/I$ can be seen if the azimuths 
differ exactly by 180$^{\circ}$. However we do not show them here, as those 
azimuths do not fall on the Carlsson grid points.

An important difference between 1D and multi-D RT is that, in the case 
of resonance scattering, multi-D geometry produces a non-zero $U/I$ 
throughout the line profile, but the 1D geometry does not produce 
any $U/I$. The Hanle scattering produces non-zero $U/I$ in 1D geometry
only in the line core, where as in multi-D RT Hanle scattering only modifies
the $U/I$ generated by the resonance scattering. Further, a strong 
$\lambda$ dependence of the spatial distribution shows that the scattering 
physics described through PRD strongly couples with spatial inhomogenity of 
the atmosphere.  

Spatial variations in the linear polarization profiles are observationally
noticed in the wings of the well known chromospheric line, the 
Ca {\sc i} 4227 \AA\, \citep[see][]{biandaetal03,sametal09}, which the 
authors refer to as the `enigmatic wing features'. 
We find such spatial inhomogeneities in the line wings of Ca {\sc ii} K
line, which is also a chromospheric line. Therefore we believe through
our studies that, the spatial variations observed in the wings of 
chromospheric lines possibly have their origin in the spatial 
structuring of the atmosphere. To understand and model 
these observations, accurate non-LTE 
multi-D RT calculations with PRD scattering are essential.
We note here that even in an isothermal atmosphere the multi-D RT 
can cause strong spatial inhomogeneities in 
the line wings through spatial structuring \citep[namely the geometry itself, 
see Figures 13 and 14 of][]{anuknn11b}. 
As it is well known, the approximation of CRD cannot explain such spatial 
structuring, because $Q/I$ and $U/I$ approach zero in the line wings, under 
this approximation. 

\subsubsection{$Q/I$ and $U/I$ in the line core}
\label{r14}
It may be interesting to notice that the linear polarization observations 
of the Ca {\sc ii} K line at 3933 \AA\, in the line core region studied 
in \citet[][]{jos06} show a strong spatial variation (in the 
range 3933.5 \AA\,--3934 \AA\,). 
Such spatial variations are caused
by (1) spatially varying magnetic fields (Hanle effect) and (2) spatial 
inhomogeneities of the atmosphere itself. These two factors are 
entangled with each other in the line core and are hard to be separated. 
Since the atmosphere chosen in this paper does not have any horizontal 
inhomogeneity in the heights where the line core is formed, we discuss only 
the magnetic field effects (Hanle effect) on the line core in this paper
(see Figures~\ref{fig-iqu1}--\ref{fig-iqu4}, particularly the middle panels).

In our studies we consider two different, spatially independent (i.e., constant)
magnetic field configurations (see Section~\ref{r1}). 
Changes in the values of $Q/I$ and $U/I$ in the line core for two magnetic
field values can be clearly observed by comparing middle panels of 
Figures~\ref{fig-iqu1} and \ref{fig-iqu3} with the corresponding 
panels of Figures~\ref{fig-iqu2} and \ref{fig-iqu4} respectively. Also one 
can observe faint spatial variation in the line core in each of these figures. 
Spatial homogeneity in the large parts of the
line core is due to the homogeneity of the chosen model atmosphere in the 
heights where the line core is formed and also because of the spatially 
independent magnetic field configurations. Considering 
a magnetic field configuration that depends explicitly on spatial 
variables $(x,z)$ and perhaps also on the angular variables $(\theta,\varphi)$ 
and/or the use of atmosphere with spatial inhomogeneity in the chromosphere 
may result in a spatially varying line core polarization similar to those 
in \citet[][]{jos06}. However we do not take up such studies in this paper.

Here one can observe that the symmetry, anti-symmetry of $Q/I$ and 
$U/I$ in the line wings (see Section~\ref{r13}) breaks in the line core region 
due to the presence of magnetic fields \citep[see][for details]{anuetal11a}.

\subsection{Comparisons with observations}
\label{compare-observations}
In \citet[][]{reneetal06,renejos07a} and 
\citet[][]{renejos07b} the authors study in detail, the linear polarization 
($Q/I$) in Ca {\sc ii} K line at 3933 \AA\, using 1D solar 
model atmospheres. Their studies suggest 
that `none of the existing 1D model atmospheres are able to reproduce the 
observations at different $\mu$ values'. 
They find that by modifying the temperature structure one
could find optimum fits to the observations. However a single model atmosphere
with a fixed temperature structure could not fit the observed $Q/I$ at 
different $\mu$ values. In other words observations of $Q/I$ at different
$\mu$ values required different modifications of the temperature structure
of the chosen 1D model atmosphere such as FALC \citep[see][]{renejos07a}. 
They conclude that use of multi-D MHD atmospheres with multi-D RT may be 
necessary to fit the observations at different $\mu$ values using a single 
model atmosphere.

In this paper we do not attempt to model the observed profiles through 
modifications of the temperature structure in the atmosphere. Our aim is
to study how well the structuring in the model atmosphere can reproduce the 
observed $(I/I_c,Q/I)$ profiles. In addition to the spatially averaged, 
emergent ($Z=Z_{\rm max}$) $(I/I_c, Q/I)$ profiles (averaged over the $X$ 
grid) we compare the observed profiles with the emergent $(I/I_c, Q/I)$ 
profiles at each of the 62 grid points along the $X$ direction and try
to fit the observations. We consider such comparisons 
because in general, spatial averaging is reasonable if the profiles do not
show strong spatial variations. However the linear polarization that we have 
obtained from the MHD-FALC atmosphere is extremely spatially inhomogeneous 
in the line wings (see Figures~\ref{fig-iqu1}--\ref{fig-iqu4}). Since each of 
these 62 points correspond to a different spatial location, the profiles at 
these points are also different. In Figure~\ref{fig-obs-theory} we plot 
observed $(I/I_c, Q/I)$ profiles (blue solid lines) with spatially 
averaged, emergent $(I/I_c, Q/I, U/I)$ profiles (red dash-triple-dotted lines) 
and spatially resolved $(I/I_c, Q/I, U/I)$ profiles (black solid lines). 

\subsubsection{The line wings}
\begin{figure*}
\centering
\includegraphics[scale=0.8]{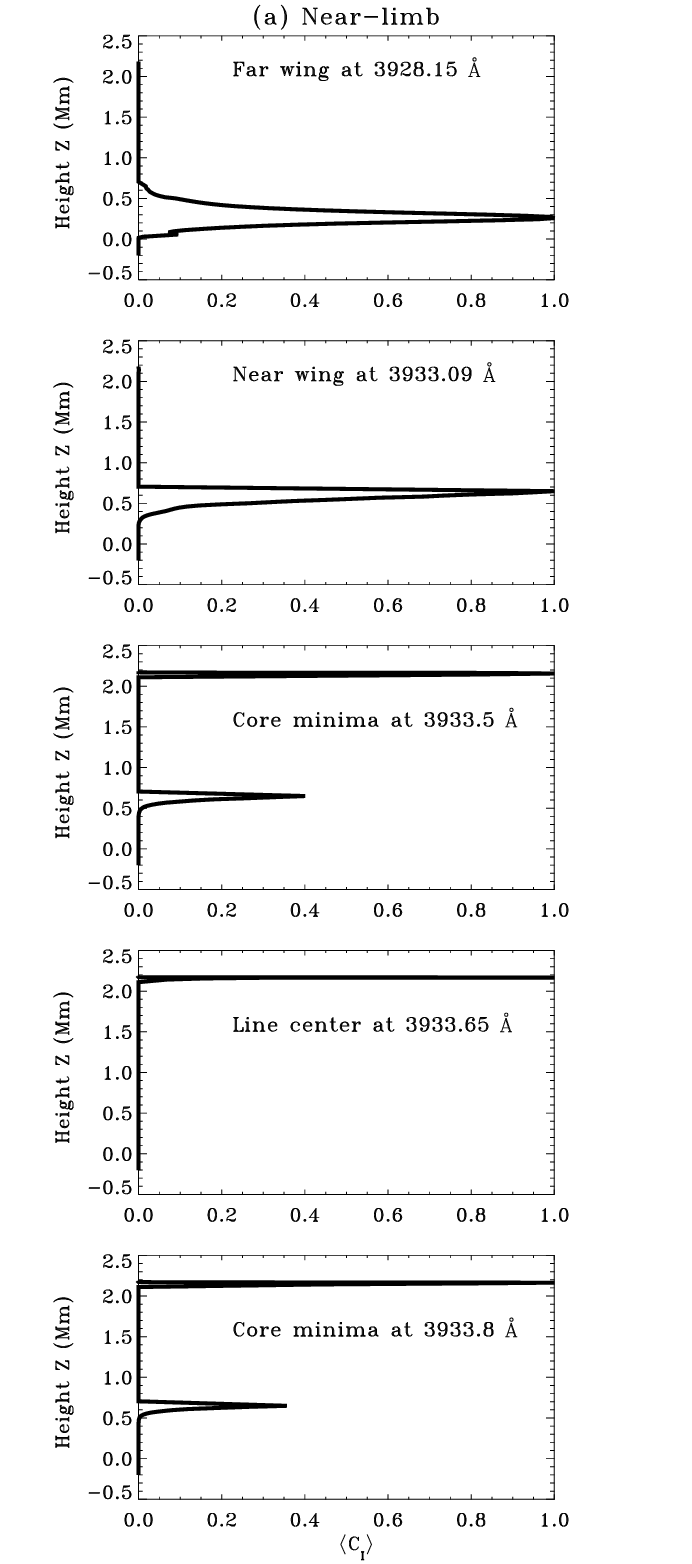}
\includegraphics[scale=0.8]{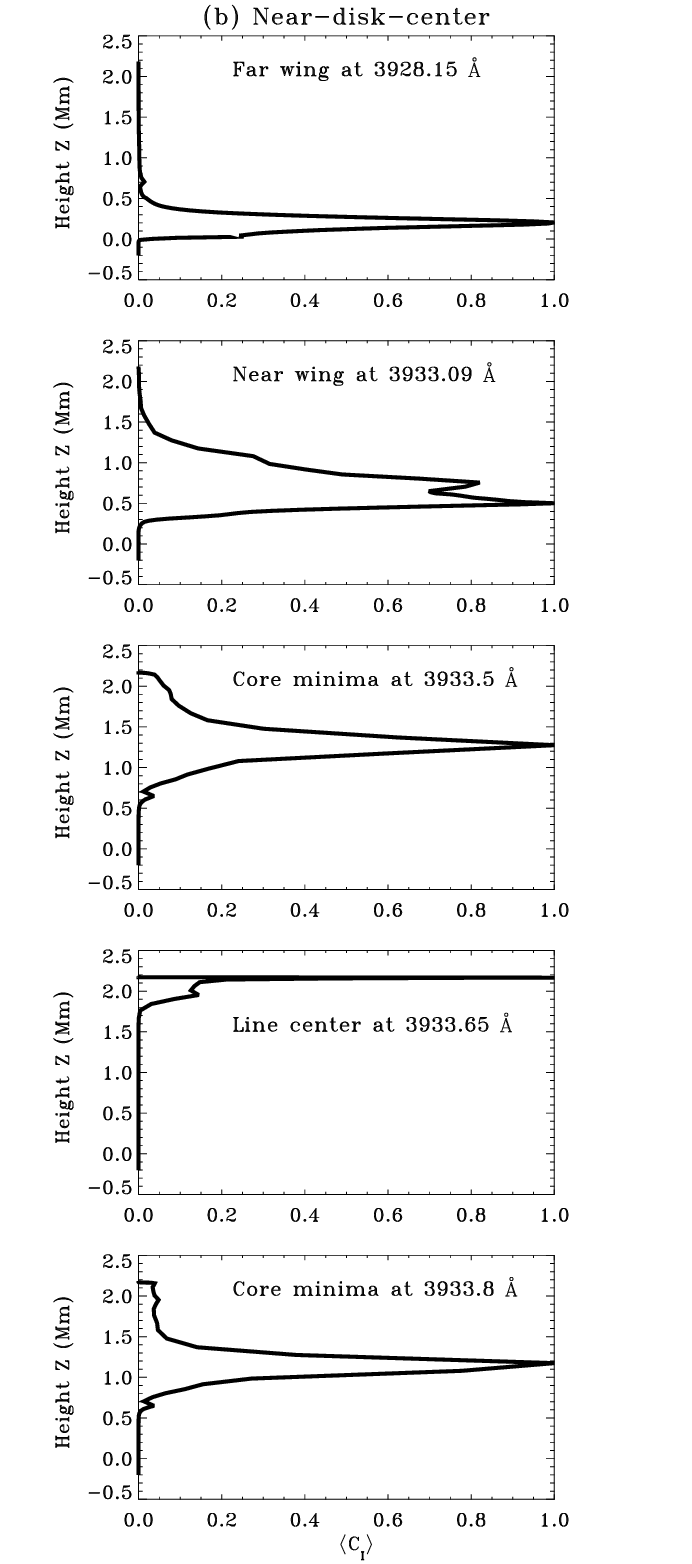}
\caption{Spatially averaged contribution function at selected wavelengths. 
Panel (a) is for near-limb direction $(\mu,\varphi)=(0.3,160^{\circ})$ and 
panel (b) for a near-disk-center direction $(\mu,\varphi)=(0.8,135^{\circ})$.
We have normalized the $<C_I>$ by the maximum value of $<C_I>$.
}
\label{fig-contrib}
\end{figure*}

\begin{figure*}
\centering
\includegraphics[scale=0.5]{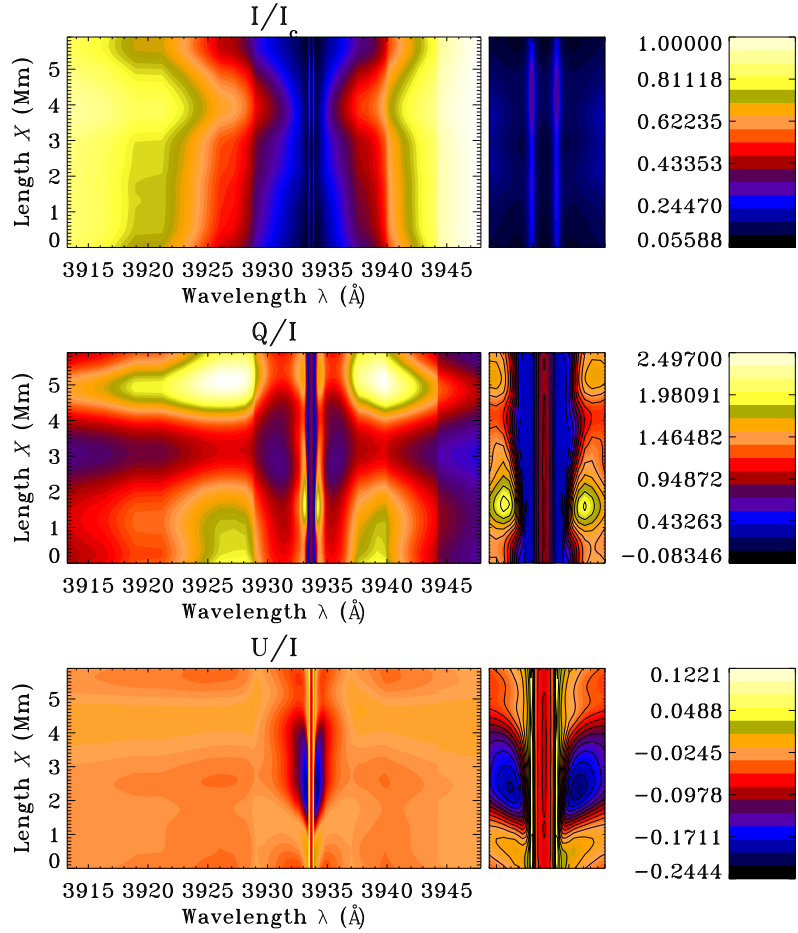}
\caption{Spatial and spectral variation of $I/I_c, Q/I, U/I$ for 
$(\mu,\varphi)=(0.3,160^{\circ})$. The spatial dimension 
($X$) represents the horizontal
variation on the surface of a 2D atmosphere. The magnetic field configuration
is $(B,\theta_B,\chi_B)=(20\,{\textrm{G\,}},45^{\circ},225^{\circ})$. 
The middle panels show the same profiles in a smaller wavelength range 
between 3932.9 \AA\, to 3934.2 \AA\,. This simulated image can be visualized 
as analogous to a spectrum obtained from an imaging spectrograph, with 
wavelength information on the horizontal axis and spatial information along 
the slit on the vertical axis.
}
\label{fig-iqu1}
\end{figure*}

\begin{figure*}
\centering
\includegraphics[scale=0.5]{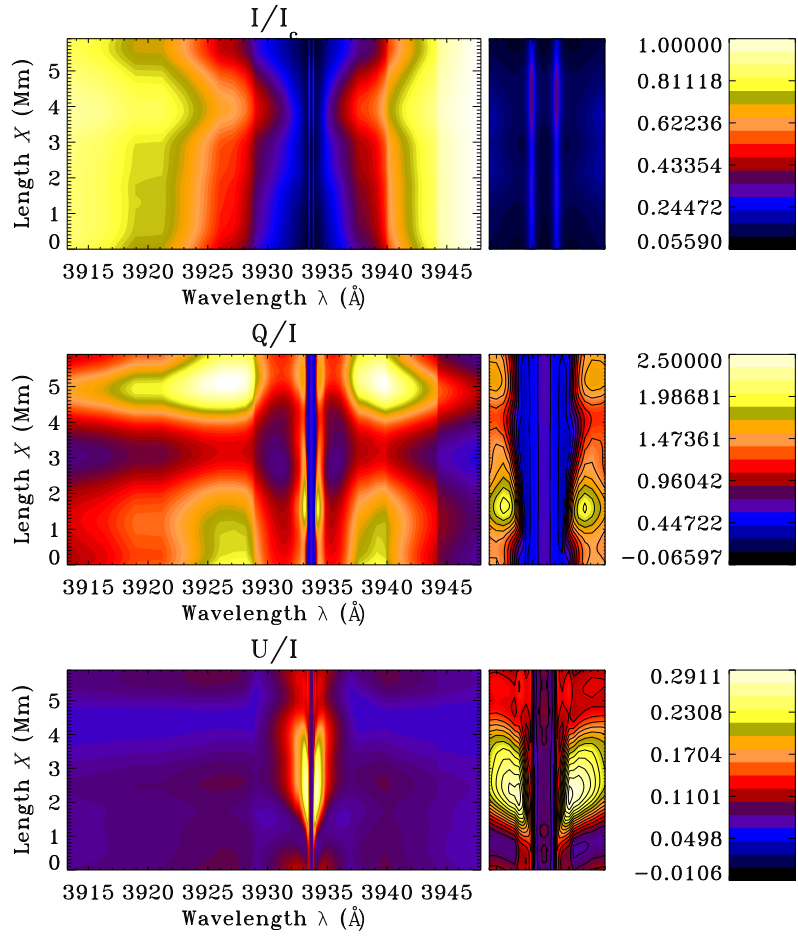}
\caption{Same as Figure~\ref{fig-iqu1}, but for 
$(\mu,\varphi)=(0.3,200^{\circ})$ and magnetic field configuration 
$(B,\theta_B,\chi_B)=(20\,{\textrm{G\,}},75^{\circ},225^{\circ})$.
}
\label{fig-iqu2}
\end{figure*}

\begin{figure*}
\centering
\includegraphics[scale=0.5]{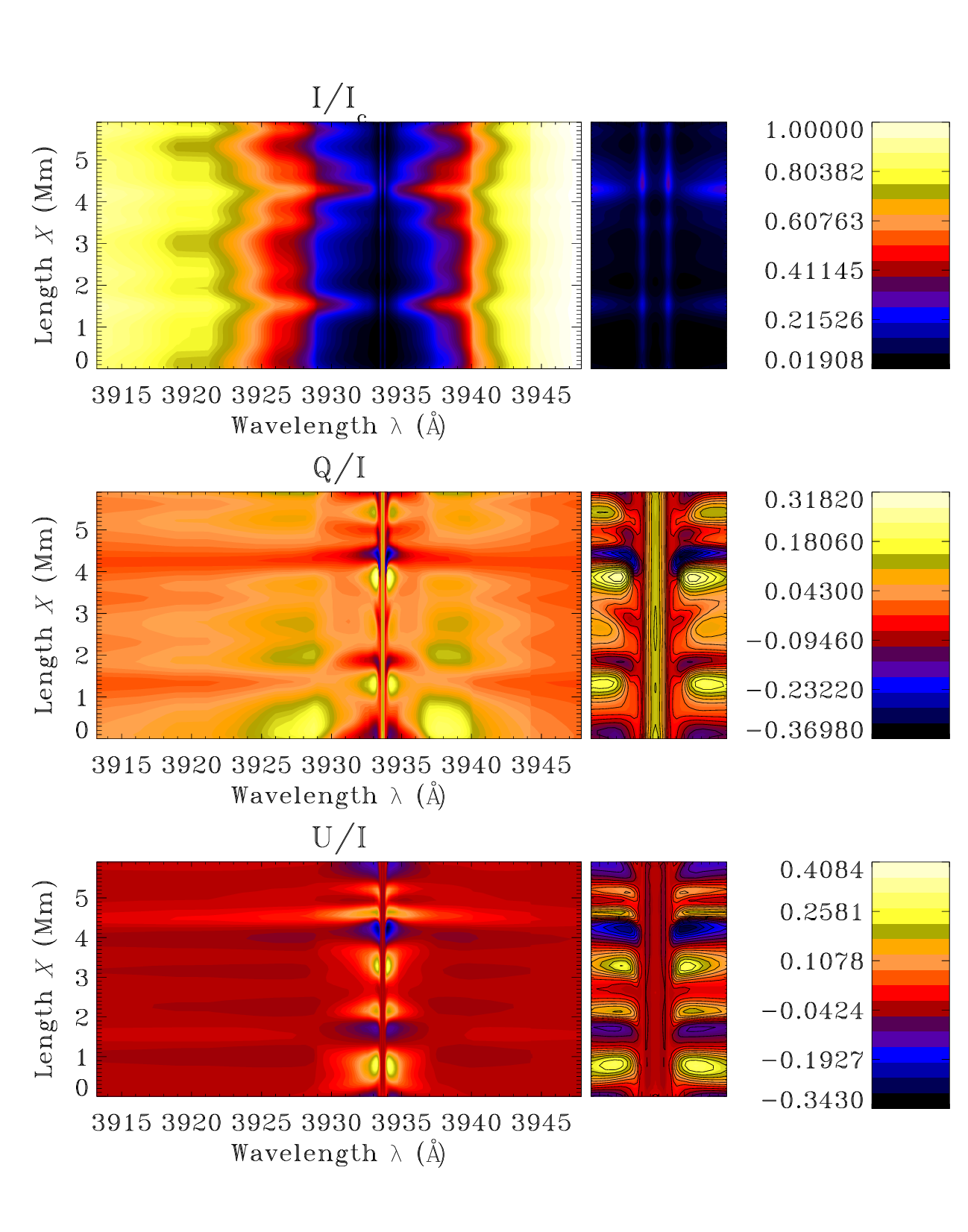}
\caption{Same as Figure~\ref{fig-iqu1}, but for 
$(\mu,\varphi)=(0.8,135^{\circ})$. 
}
\label{fig-iqu3}
\end{figure*}

\begin{figure*}
\centering
\includegraphics[scale=0.5]{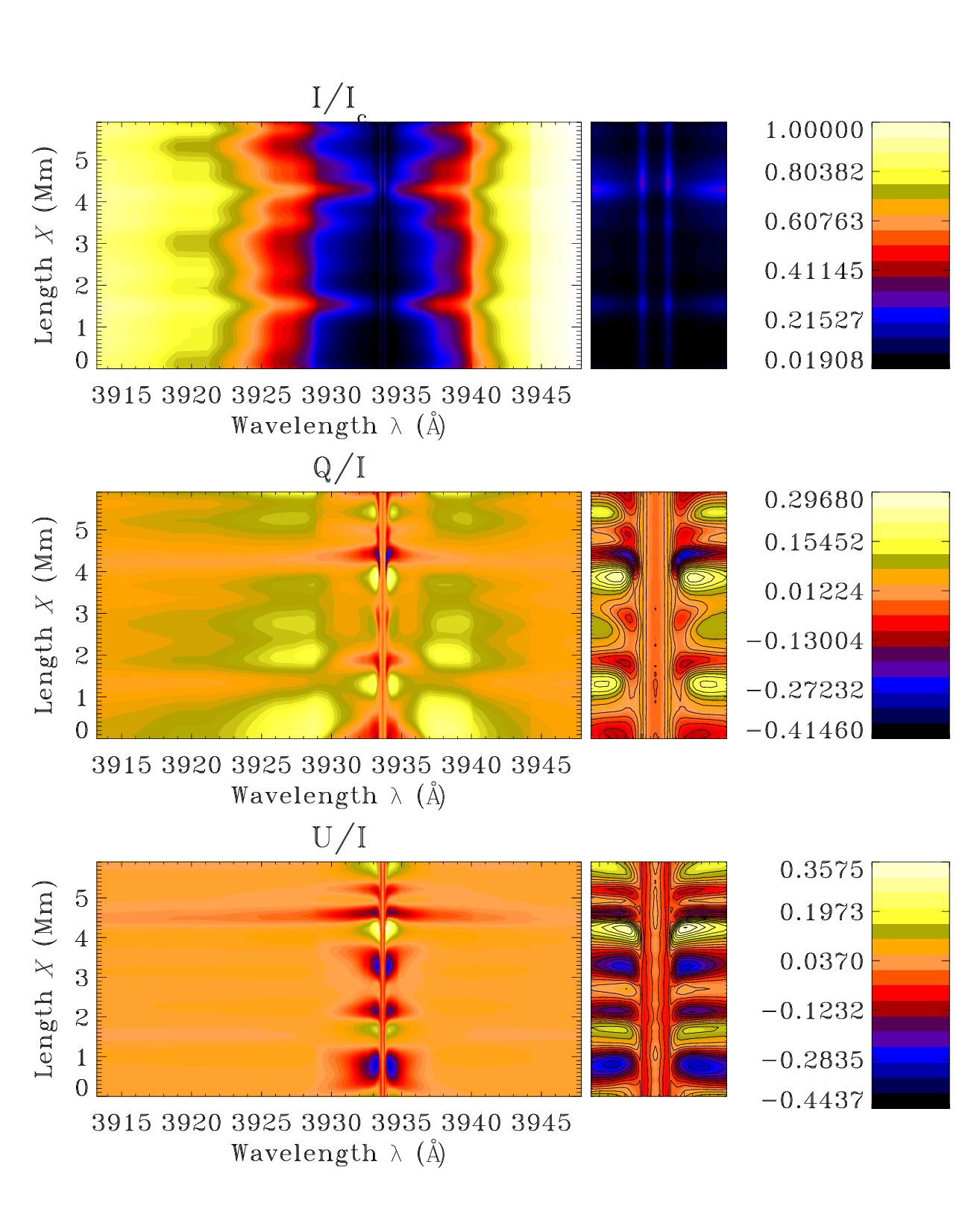}
\caption{Same as Figure~\ref{fig-iqu2} but for 
$(\mu,\varphi)=(0.8,225^{\circ})$.
}
\label{fig-iqu4}
\end{figure*}

\begin{figure*}
\parbox{17cm}{
\includegraphics[scale=0.8]{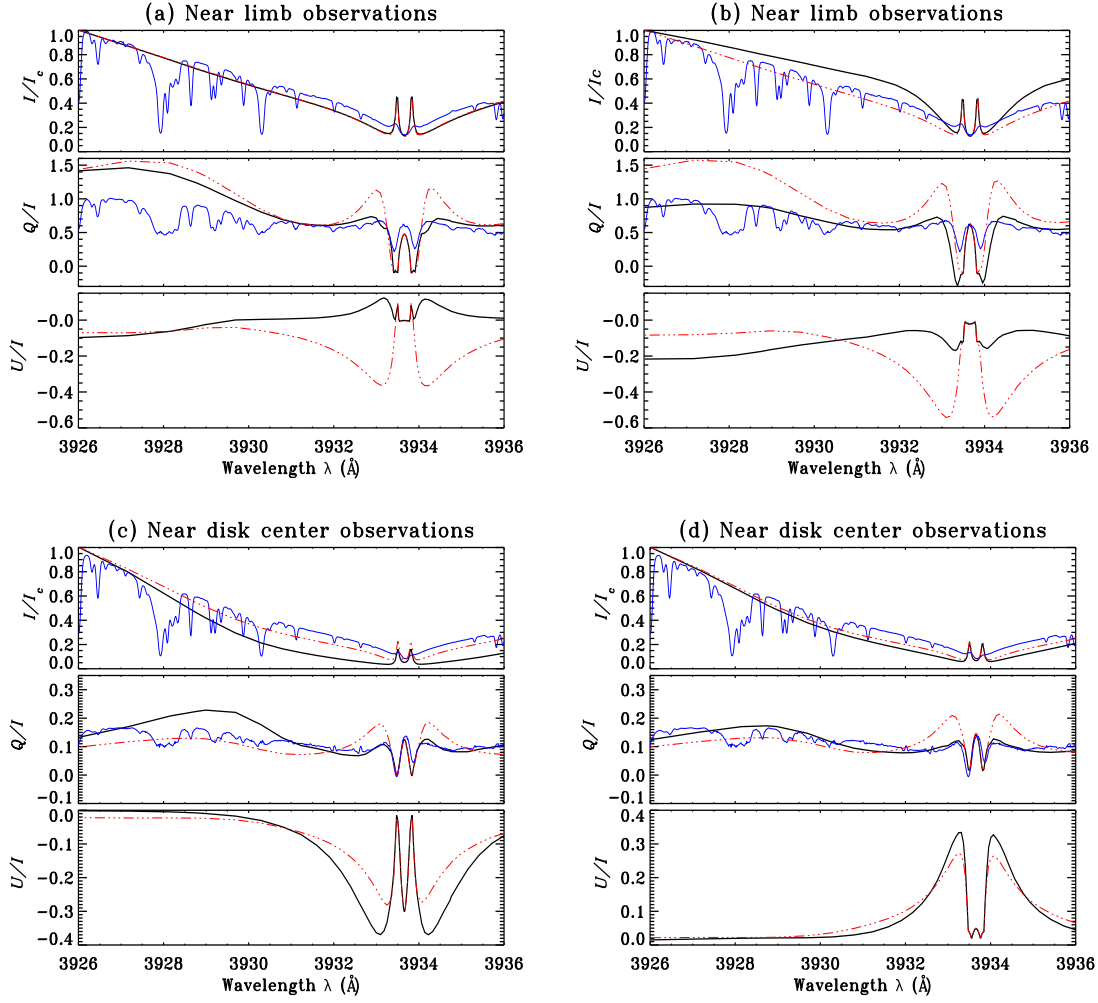}
\caption{Comparison of $(I/I_c, Q/I)$ observations of the Ca {\sc ii} K
line with the theoretical $(I/I_c, Q/I, U/I)$ profiles. 
We show observations (blue solid lines), spatially averaged theoretical 
profiles (red dash-triple-dotted lines) and spatially resolved theoretical 
profiles (black solid lines). The ray directions, spatial locations (for 
spatially resolved curves only) and the 
magnetic field configurations considered in panels (a)--(d) are respectively: 
(a): $(\mu,\varphi)=(0.3,111^{\circ})$, $X=$0.45 Mm, 
$\bm{B}=(20\,{\rm G\,}, 90^{\circ}, 225^{\circ})$, 
(b): $(\mu,\varphi)=(0.3,70^{\circ})$, $X=$0.51 Mm, 
$\bm{B}=(13\,{\rm G\,}, 90^{\circ}, 225^{\circ})$ 
(c): $(\mu,\varphi)=(0.8,45^{\circ})$, $X=$0.1 Mm, 
$\bm{B}=(20\,{\rm G\,}, 45^{\circ}, 45^{\circ})$  
and (d): $(\mu,\varphi)=(0.8,315^{\circ})$, $X=$0.54 Mm, 
$\bm{B}=(60\,{\rm G\,}, 120^{\circ}, 45^{\circ})$.}
\label{fig-obs-theory}
}
\end{figure*}

The MHD-FALC atmosphere chosen in this paper has spatial
inhomogeneities only up to a height of $\sim$ 0.65 Mm. Since the line wings
(defined in Section~\ref{r1}) are formed at or below this height (see 
Section~\ref{contrib} and Table~\ref{table_1}), we expect that the 
line wings should be well reproduced by the MHD-FALC atmosphere. 
In Figure~\ref{fig-obs-theory} we can 
see that better fit to the envelop of the wings of the $I/I_c$ profiles could 
be obtained with the spatially averaged profiles than with the spatially 
resolved ones. However for the $Q/I$, 
the spatially averaged profiles could not fit the observations at all, perhaps
because the wing polarization is extremely spatially inhomogeneous.
We could reasonably fit the line wings only by using the spatially resolved 
profiles. In Figure~\ref{fig-obs-theory} (a)--(d) the profiles are plotted 
for the spatial locations $X=$ 0.45 Mm, 0.51 Mm, 0.1 Mm and 0.54 Mm 
respectively. 
In the case of near-limb observations, we could not obtain a simultaneous
fit to the far and the near wings at any of the 62 spatial locations. 
Figure~\ref{fig-obs-theory}(a) is an example of a reasonable fit to the near 
wings and Figure~\ref{fig-obs-theory}(b) for the far wings. 
Since the far wings are formed well below 0.65 Mm, we expect the MHD variation
in this lower atmosphere to well reproduce them. The near wings are difficult 
to reproduce using MHD-FALC atmosphere because, 
the near wing region which we are unable to fit in 
Figure~\ref{fig-obs-theory}(b) is formed just around the height $\sim 0.65$ Mm 
(see Figure~\ref{fig-contrib}(a)) which is the height at which the atmosphere 
changes from 2D-MHD variation to 1D FALC, and thus around these heights, 
the atmosphere may have some deficiencies. 
In the case of near-disk-center observations,
Figure~\ref{fig-obs-theory}(d) represents a better fit to the line wings
(both near and far) than Figure~\ref{fig-obs-theory}(c). We are able to 
reasonably fit the near-wings in this case because they are formed around 
$\sim 0.5$ Mm (see Figure~\ref{fig-contrib}(b)) where the atmosphere is well
represented by MHD variations.

\subsubsection{The line core}
The line core is formed above a height of 0.65 Mm (see Section~\ref{contrib})
where the MHD-FALC atmosphere does not have any horizontal inhomogeneities.
Therefore we do not expect the atmosphere to provide a good fit to the 
observations. We see that in Figure~\ref{fig-obs-theory} even the 
$I/I_c$ profiles are not well fitted. 
However by computing the theoretical profiles for several magnetic field 
configurations, it is possible to obtain a good fit to the line center 
value of $Q/I$. The optimum values of the magnetic fields for obtaining the fit 
in Figure~\ref{fig-obs-theory} are: 
$\bm{B}=(20\,{\rm G\,}, 90^{\circ}, 225^{\circ})$ for panel (a), 
$\bm{B}=(13\,{\rm G\,}, 90^{\circ}, 225^{\circ})$ 
for panel (b), $\bm{B}=(20\,{\rm G\,}, 45^{\circ}, 45^{\circ})$ for panel (c) 
and $\bm{B}=(60\,{\rm G\,}, 120^{\circ}, 45^{\circ})$ for panel (d).
The choice of different magnetic fields helped to obtain a reasonable fit to the
line core $Q/I$ in the case of near-disk-center observations. We could not 
reproduce the line core of $Q/I$ (except the line center) for near-limb
observations for any choice of the magnetic field values that we considered.
The fact that 1D models can reproduce the observations of the line core of the 
near-disk-center but not those of the near-limb is also true for
the Ca {\sc i} 4227 \AA\, line, which we found in our previous
studies \citep[see][]{anuetal10,anuetal11b}. This would perhaps 
indicate that the heights at which line cores
are formed in the case of near-limb observations are more inhomogeneous than
the heights where the line core is formed in the case of the near-disk-center 
observations (see Section~\ref{contrib}, Figure~\ref{fig-contrib} and 
Table~\ref{table_1}). However, we cannot attribute it to the goodness of the
atmosphere because a consistent atmosphere will have inhomogeneities 
at all the heights and it should be able to reproduce the entire line profile
at all the lines of sight.

\section{CONCLUSIONS}
\label{conclusions}
This paper represents the first application of our work in a previous series of 
papers on polarized RT in multi-D media. Here we develop the necessary code,
test it and use it to study the linear polarization in the Ca {\sc ii} K line
at 3933 \AA\,. This work is an initial step towards modeling the chromospheric 
lines because here we choose an approximate model atmosphere. It is a 2D 
snapshot of an MHD atmosphere in the photosphere combined with columns of 1D 
FALC atmosphere in the chromosphere. The use of PRD as the line scattering 
mechanism is essential in the modeling effort, because the approximation of CRD 
leads to nearly zero linear polarization in the line wings. 

The MHD structuring in the photosphere produces spatial inhomogeneities
in the wings of the linear polarization profiles and it explains the causes of
spatial structuring observed in the wings of some strong chromospheric 
lines.

Since the atmosphere that we have used can represent the MHD structuring up to
a height of $0.65$ Mm, the wavelength region of the $Q/I$ profiles formed below
this height could be fitted reasonably well. This includes far wings for both
the lines of sight (near limb and near-disk-center), near wings in the case
of near-disk-center observations. Using different magnetic field 
configurations and Hanle effect we obtained reasonable fits to the line 
center values of $Q/I$ for both the lines of sight.
In the case of near-limb observations, the near wings and the line core 
region including core minima could not be fitted using this atmosphere. 
Although we could fit the line core region in the case of near-disk-center 
observations, the reason for a good fit cannot attributed to the 
goodness of the atmosphere because the same atmosphere could not produce the 
line core in the case of near-limb observations.

This study clearly indicates that as in the photosphere, MHD structuring in the
chromosphere is an important requirement to obtain simultaneous fit to the
line core and the line wing polarization observations of the chromospheric 
lines at all the lines of sight. To achieve this, we need 3D MHD model 
atmospheres that can quite well represent the solar chromospheric 
inhomogeneities.
\appendix
\section{Pre-BiCG-STAB algorithm}
\label{prebicg}
In this Appendix we describe some important steps in the Pre-BiCG-STAB 
algorithm, to be taken care when using the method to model 
spectral lines, for which both line and continuum scattering processes become 
important. 

Using the formal solution expression for $\bm{\mathcal{I}}$,
the vector $\bm{\mathcal{J}}$ in Equation~(\ref{jbar})
can be written as
\begin{eqnarray}
&&\bm{\mathcal{J}}(\lambda,\bm{r})=
\Lambda[\bm{\mathcal{S}}(\lambda,\bm{r})].
\label{lambda}
\end{eqnarray}
Similarly the continuum source vector in Equation~(\ref{sc-reduced}) can 
be written as
\begin{eqnarray}
&&{\bm{\mathcal S}}_{c}(\lambda, \bm{r})=
\Lambda_c[\bm{\mathcal{S}}(\lambda, \bm{r})].
\label{lambda-c}
\end{eqnarray}
Recall the total source vector expression given by 
\begin{equation}
\bm{\mathcal S}(\bm{r}, x)=p_l[\epsilon {B}_{\lambda}(\bm{r})\bm{\mathcal U}+
\bm{\mathcal J}(\bm{r}, x)]
+ p_c\bm{\mathcal S}_c(\lambda,\bm{r}) + p_a {B}_{\lambda}(\bm{r}) 
\bm{\mathcal U}.
\label{tot-s}
\end{equation}
Substituting Equations~(\ref{lambda}) and (\ref{lambda-c}) in 
Equation~(\ref{tot-s}), we obtain a system of equations
\begin{eqnarray}
&&\hat{A} \bm{\mathcal S}=\bm{b}\quad \quad \textrm{or}\quad \quad
[\hat{I}-p_l\Lambda-p_c\Lambda_c]\bm{\mathcal S}(\lambda, \bm{r})=
[p_l\epsilon + p_a]{B}_{\lambda}(\bm{r})\bm{\mathcal U}.
\label{sys}
\end{eqnarray}
Let $\bm{\mathcal{S}}_0$ denote an initial guess for the source vector.
In this paper we calculate the initial source vector using the unpolarized 
mean intensity obtained by solving the unpolarized multi-level RT equation
in the RH-code. The rest of the algorithm is similar to that described in 
\citet[][]{anuetal11a}.

\acknowledgements
We would like to thank Dr. Han Uitenbroek for useful discussions and for 
providing a version of his RH-code. We thank him also for providing the 
MHD-FALC atmosphere used in this paper. 
We are grateful to Dr. Dominique Fluri and Dr. Rene Holzreuter who kindly
provided their 1D code for calculating the 1D results. 
We would like to thank Dr. Baba Varghese, 
IIA, for his kind help in generating some of the idl plots. 
L. S. A. would like to thank Dr. Michiel van Noort for very 
useful discussions. We thank Dr. Rene Holzreuter for providing the observed
data shown in this paper. We thank J. O. Stenflo, A. Feller, C. Thalmann, 
D. Gisler, M. Bianda, R. Ramelli, A. Gandorfer, and C. U. Keller 
who performed these observations.


\end{document}